# IoT Book Bot

Souvik Datta
*School of Electrical Engineering*
*Vellore Institute of Technology*
Chennai, India
souvik.datta2019@vitstudent.ac.in

Mangolik Kundu
*School of Electrical Engineering*
*Vellore Institute of Technology*
Chennai, India
mangolik.kundu2019@vitstudent.ac.in

Ratnadeep Das Choudhury
*School of Electrical Engineering*
*Vellore Institute of Technology*
Chennai, India
ratnadeepdas.choudhury2019@vitstudent.ac.in

Dr. Sriramalakshmi P
*School of Electrical Engineering*
*Vellore Institute of Technology*
Chennai, India
sriramalakshmi.p@vit.ac.in

Dr. Sreedevi V T
*School of Electrical Engineering*
*Vellore Institute of Technology*
Chennai, India
sreedevi.vt@vit.ac.in

*Abstract*— **In order to ease the process of library management many technologies have been adopted but most of them focus on inventory management. There has hardly been any progress of automation in the field of issuing and returning books to the library on time. In colleges and schools, hostellers often forget to timely return the issued books back to the library. To solve the above issue and to ensure timely submission of the issued books, this work develops a 'Book-Bot' which solves these complexities. The bot can commute from point A to point B, scan and verify QR Codes and Barcodes. The bot will have a certain payload capacity for carrying books. The QR code and Barcode scanning will be enabled by a Pi Camera, OpenCV and Raspberry Pi, thus making the exchange of books safe and secure. The odometry maneuvers of the bot will be controlled manually via a Blynk App. This paper focuses on how human intervention can be reduced and automates the issue part of library management system with the help of a bot.**

*Keywords—Raspberry Pi, Book Bot, Arduino UNO, Blynk, HX-711 Load Sensor.*

## I. INTRODUCTION

Even though a major part of modern-day library management system has been automated still there remains a significant gap in the field of timely and secured submission of books back to the library. As per a survey conducted on sixty-one randomly selected medical students from the University of Ilorin, it was found that most of the overdue books were a result of either students being forgetful or the students not finishing with their books. It was also found that renewal was the most favored overdue measure among the students [1]. Similarly, in a survey conducted by Shontz in 1997 [2], his studies indicated that factors such as convenience of renewal and returns, and time when the user has finished with the books, were more important in determining when users could return library materials than imposing fines. From almost all of these studies, it is evident that forgetfulness is the primary reason of overdue books and late submissions. Thus, to overcome the above issue and to ensure ease and timely submission of issued books to the library, this work suggests a robot as a solution.

This Book-Bot which is essentially a robot carrying books will help reduce the workload of the library management system. Moreover, it will also save time and make it a much more hassle-free process for students. It will be enabled with a barcode scanner to ensure the correct submission of books and will also have a QR code scanner that will scan a unique QR code provided to every student to verify the student details. After submission, it will store the submission information in the database. After the payload capacity of the robot is full, a staff member can remotely control and drive the robot to the library to submit the books. To ensure more safety for pedestrians, the robot will be enabled with Ultrasonic sensors and GPS based tracking as well.

In a campus, the Book-Bot can be placed at the Hostel, which will remind students about their overdue submissions, if any. Once students come to submit their books, they will first be asked to scan their individual unique QR Codes to verify their details. Following which, the bot will be able to access the database and find out data regarding their pending or upcoming submissions. Following this, the students will be shown a list of books that they have borrowed from the library, from the list they will be required to select the books that they wish to submit. After this, for each book, they will be given an option of whether they wish for a renewal or if they wish to submit the book. If the student wishes for renewal the book will be renewed for one week and the same gets updated in the database as well, whereas in order to submit the book, the student would be required to scan the Barcode of the specific book and submit it to the Book-Bot. Once the payload capacity of the robot is full, a staff member can remotely control and drive the robot back to the library.

## II. RELATED WORKS

Previous works have explored the development of various library automation systems which primarily operate in closed and controlled environments. Moreover, a significant amount of research has been dedicated to robotic arms in assisting library management systems [3]. Few research projects have also tried to integrate similar technologies using Raspberry Pi for semi-autonomous bots [4][5][6], but they lack feedback-based motor control, hence they cannot be used in real life applications. Attempts have been made to automate library services using Near Field Communication (NFC) embedded tags and Radio Frequency Identification (RFID) technologies, however implementing such technologies involve heavy investments [7][8]. Moreover, certain robots using these technologies [9] cannot be tracked or transversed to places beyond the pre-defined RFID boundary. While some ideas include the use of RFID tags and line following robot to eliminate issues regarding misplacement of books others include the use of aerial drones and computer vision to address the same [10][11]. While almost all of these ideas are designed to work with well demarcated indoor areas within the library none of them are designed to work with outdoor environments. They don't solve issues of addressing overdue submissions or keeping track of re-issued books. The idea of automating the process of remote outdoor book collection and returning them back to the library isn't discussed in these literatures.

Certain developments have been made using Bluetooth to track and locate warehouse assets [12]. Unfortunately, Bluetooth provides an extremely short range of connectivity and is thus not feasible for any outdoor deployment. To the best of our knowledge, no work/technology has managed to seamlessly ease the process of automation or lacks a robust security framework. Furthermore, RFID readers are 10 times more expensive than barcode readers; this research project removes the entire cost by implementing OpenCV libraries. As discussed below, the proposed IoT Book Bot achieves greater robustness while increasing reproducibility.

### III. METHODOLOGY

The Bot encapsulates the use of Raspberry Pi and Pi Camera for all computer vision operations. The bot is designed to transverse via a remote control while returning a live video stream and Global Positioning System (GPS) location. Furthermore, it also provides a secure interface for book exchange. This will ensure the timely submission of books in a hassle-free manner. The technological aspects of the Bot are described in the following sections:

*A. QR Code and Barcode Decryption*

Our proposed model uses Python's ***pyzbar module*** to decode one dimensional QR-Code of students and barcodes from Books to update the database during submissions. Using the 'decode' function of the library, which returns an array of objects, with each element of the array representing the detected QR/Barcode. The `decode()` returns a decoded tuple consisting of these four –

1. `data` – A strings in bytes which needs to be further decoded using **utf-8** to return a string.
2. `type` – Contains the type of decoded barcode
3. `rect` – Represents the captured localized area
4. `polygon` – List of point instances representing the barcode.

*B. Blynk Communication Interface*

Blynk was designed to control hardware, access and visualize sensory data in real-time over IoT. It consists of a full suite of software required to prototype, deploy and remotely manage connected devices. Messages sent over Blynk Cloud is encrypted with Transport Layer Security (TLS) protocol. In this paper, Blynk is integrated with a NodeMCU ESP-8266 for motor control the data is also visualized from the HC-SR04 Ultrasonic Distance Sensor and Neo-6m GPS Module for real-time tracking.

*C. HX-711 Load Cell Amplifier Module*

HX-711 is a precision 24-bit analog to-digital converter (ADC) designed for weigh scales. HX-711 Load Cell Module converts the pressure (force) applied into an electrical signal. It is an electronic transducer that converts force into a weak micro-voltage level electric signal. It consists of four strain gauges that are arranged in a Wheatstone Bridge fashion. The strain gauges are resistors that change their resistance when bent on the application of pressure. The Amplifier Module is designed to amplify these signals and to report them to the Arduino UNO. It requires a minimal 5V supply for operating.

*D. Neo-6M GPS Module*

GPS receivers try and figure out the distance from a set number of satellites, which they locate from their pre-programming. In a process known as ***Trilateration***, it can pin point the location on Earth [13]. In this process, satellites transmit their position and current time through radio waves, while receivers here on Earth try and calculate the distance from each satellite based on the time period of the signal. A receiver needs a minimum of three such satellites to pin point its location on Earth.

The Neo-6M GPS Module covers over fifty channels, and has a sensitivity of -161 decibels(dB) while consuming only 45mA of supply current. It can also track over 22 satellites

### IV. COMPONENTS USED

*A. Component List:* The various components used during the design and prototype development of the Book Bot are mentioned in Table 1. A brief description of each component is detailed in the subsequent sections.

Table 1. Components used during Prototype Development

| Sr. No | Component Name | Unit |
|---|---|---|
| 1 | Raspberry Pi 4B | 1No |
| 2 | 5 MP Raspberry Pi Camera | 1No |
| 3 | Arduino UNO | 1No |
| 4 | HX-711 Load Cell & Amplifier Module | 1No |
| 5 | HC-SR04 Ultrasonic Sensor | 1No |
| 6 | Servo Motor | 1No |
| 7 | L298N Motor Driver | 1No |
| 8 | NodeMCU ESP-8266 Wi-Fi Module | 1No |
| 9 | High Torque DC Motors | 2Nos |
| 10 | Chassis for Bot-Platform | 1No |
| 11 | Neo-6M GPS Module | 1No |
| 12 | 11.1 V Lithium Polymer Battery (2200 mAH) | 1No |
| 13 | Jumper Wires |  |
| 14 | 16 GB Micro-SD Memory Card | 1No |

*B. Brief Description of major components:*

*1) Raspberry Pi 4B (4 GB RAM):* The Raspberry Pi is a credit card-sized computer which employs a System on a Chip (SoC). It combines a CPU and a GPU into a single integrated circuit, with a RAM, USB ports, and other components soldered to the board for an all-in-one solution. The Raspberry Pi Camera Board connects directly to the Raspberry Pi's CSI connection. It can capture images with a 5MP resolution. A Representational picture of the Raspberry Pi 4B is shown in Figure 1.

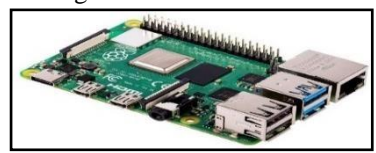

Figure 1. Representational Picture of Raspberry Pi 4B

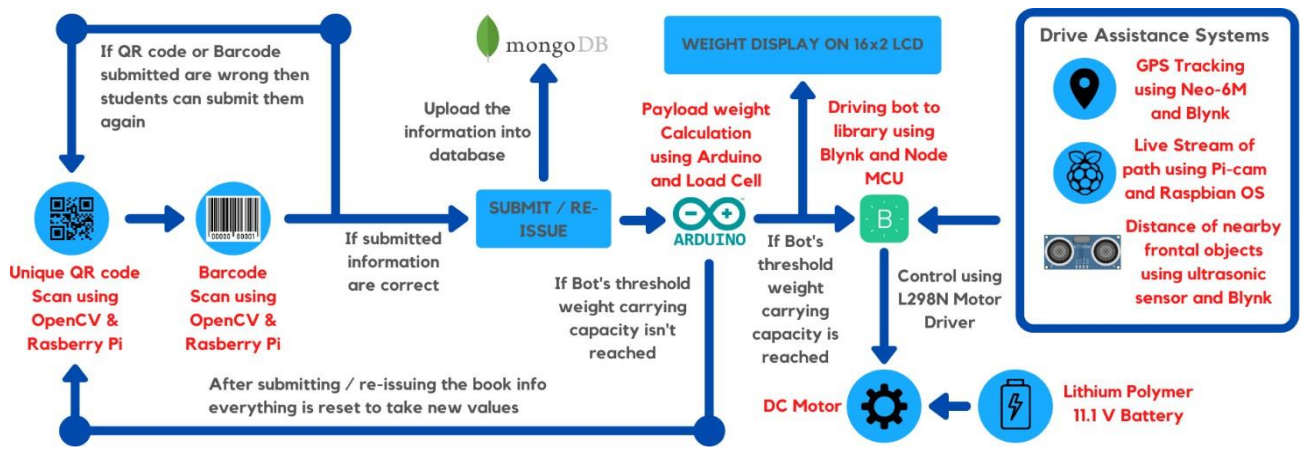

Figure 2. Block Diagram depicting the entire operation of Book Bot

The developed working of the IoT Book Bot is represented as a block diagram in Figure 2.

*2) HX-711 Load Sensor with Arduino UNO:* A strain gauge is subjected to a load, causing it to strain a particular amount and output a voltage proportionate to the applied load. This strain-voltage connection is employed in a variety of applications where weight measurement is critical. Because of their linearity, cost-effectiveness, and ease of installation, load cells are widely used. The load-cell is connected with Arduino UNO for the calculation of weight using C++. Figure 3 represents the interfacing of Arduino UNO with the HX-711 Load Sensor.

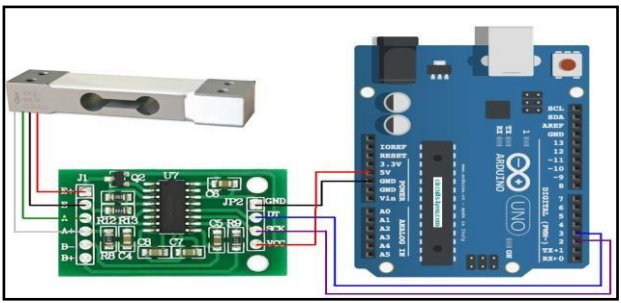

Figure 3. Arduino Uno with Load cell & Amplifier module

*3) L298N Motor Driver:* The L298N is a twin H-bridge chip, which is placed on this handy breakout board along with all essential peripherals, can drive a 2-phase bipolar stepper motor or two DC motors. The L298N Motor Driver is shown in Figure 4. It is highly suited for robotic applications and can be connected to a microcontroller with only a few control lines per motor.

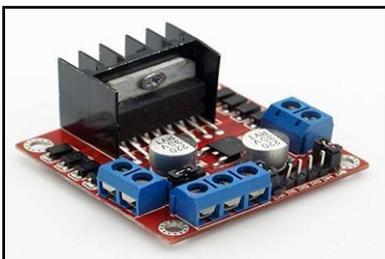

Figure 4. Representational Picture of L298N Motor Drive

*4) NodeMCU ESP8266:* NodeMCU is an open-source platform built for Internet of Things (IoT) based applications. At the heart of it is the ESP-8266, Wi-Fi chip with TCP/IP protocol [14]. It uses an on-board flash-based Serial Peripheral Interface Flash File System (SPIFFS). It has an operating voltage range of 2.5V~3.6V and draws 80mA of current during Operation Mode and 20µA during Sleep Mode. The ESP8266 combines an 802.11b/g/n HT40 Wi-Fi transceiver, this way it can not only connect to a Wi-Fi network and interact with the Internet, but also set up a network of its own, allowing other devices to connect directly to it. The representational picture of NodeMCU ESP-8266 is shown in Figure 5.

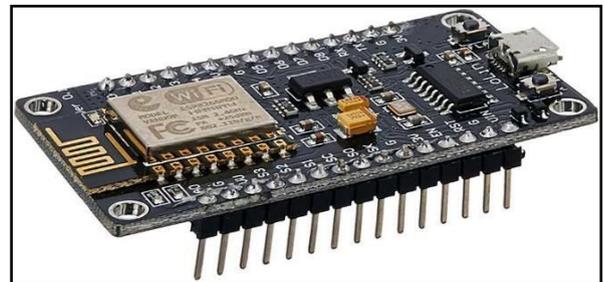

Figure 5. Representational Picture of NodeMCU ESP8266

V. HARDWARE & DISCUSSIONS OF RESULTS

The aim of this proposed system is to provide a solution to the problem of the timely return of the issued books to the library. When students wish to submit/re-issue a book then they can go near the bot and scan the barcode of the issued book by showing the same in front of the camera. Then the unique QR Code of a particular student can be scanned into the system using the same method as shown in Figure 6. On completion of the above two steps, students can select whether they wish to submit or re-issue the book by clicking the respective button on the console of the bot which is displayed on a website, hosted on an administrator's laptop.

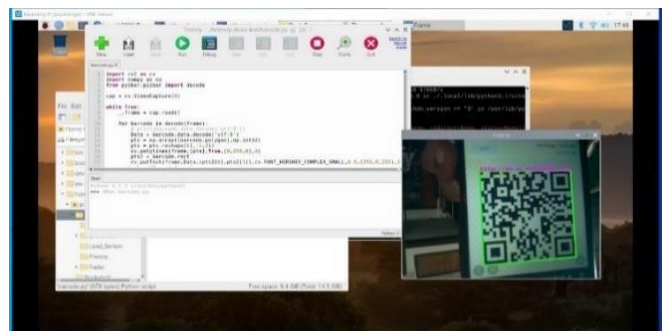

Figure 6. Unique QR code of the student being scanned

The complete circuit diagram of the proposed hardware is presented in Figure 7. Now, once the QR code of the student and the barcode of the issued book is scanned, the system uploads the above details in a database and marks the book as 'submitted' or 're-issued'. In this research project, MongoDB database has been used to achieve this functionality as shown in Figure 8.

are present on the interface of the Blynk app itself as shown in the Figure. 10 and Figure. 11. The map will fetch the real time location coordinates from a Neo-6M GPS Module.

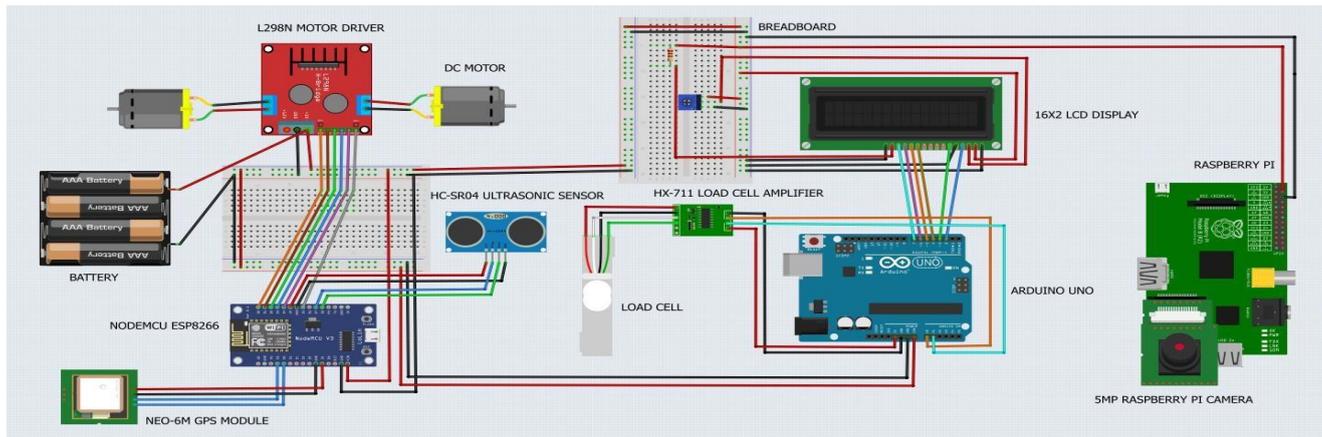

Figure 7. Circuit Diagram of proposed hardware

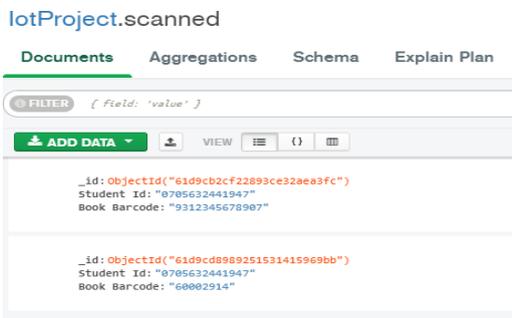

Figure 8. MongoDB Database after submission of two books with respective decoded QR code and Barcode of the book

The bot provides a system to detect weight which consists of a load cell, Arduino with code written in C++ to calculate weights, 16 x 2 LCD display to display weight and an arrangement of cards to mimic the functionality of a cantilever for weight detection as shown in Figure 9. Using this functionality, the bot will get to know the total weight of books in its inventory. If the weight exceeds the threshold amount of weight a bot can carry, then the bot stops accepting any more books and shows a warning for reaching its threshold weight. For this project, the threshold weight is 5 kgs.

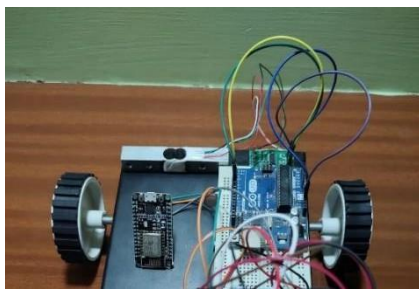

Figure 9. Load cell with Arduino UNO

When the inventory gets filled up to the threshold weight, a person can drive the bot to the library for the submission of the collected books. To drive and navigate the bot, the person can use the Blynk app where the controls for cardinal directions are present. To assist driving, a digital LCD display on which the distance from the nearby frontal object can be displayed and a map to show the real time position of the bot;

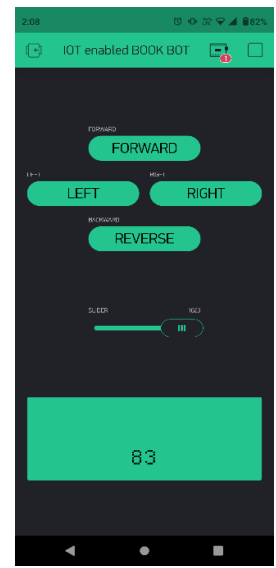

Figure 10. Blynk App Interface for Bot control

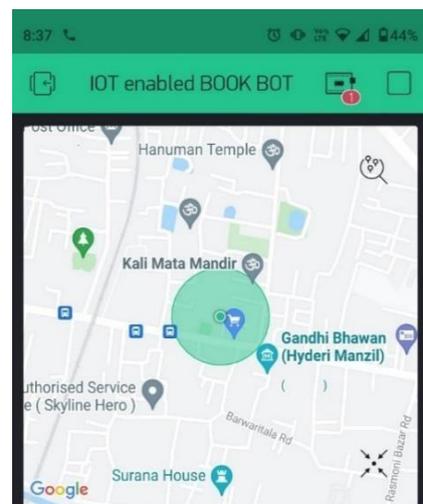

Figure 11. Blynk App Interface with Live GPS Tracking for bot control

The bot is also enabled with live stream of the path on which the bot is travelling. The live stream will be shown on a website hosted on the controller's laptop. The complete model of the IoT Book Bot is presented in Figure 12.

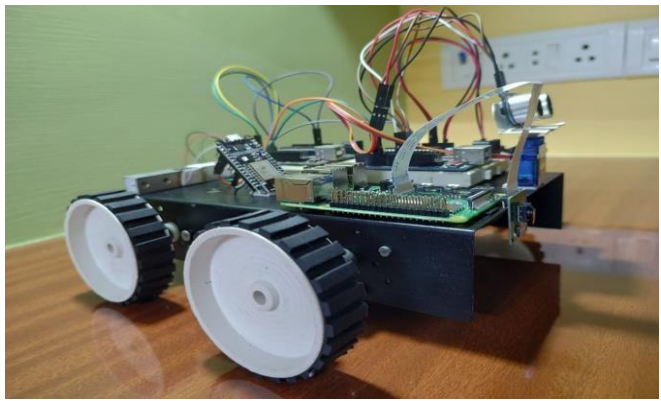

Figure 12. The IoT Book Bot with all its components

## V. Future Prospects

As the world is abuzz with semi-autonomous and fully autonomous vehicles, the most desirable objective would be to create a fully autonomous Book Bot. In view of the ongoing pandemic, peer engagement in libraries has been restricted to the minimum. Moreover, students find it difficult to access the library's resources [15]. Thus, a complete autonomous system can solve a major part of the entire logistics operation. The paper can be further upgraded to allow request-based book delivery to students.

This will require us to integrate the present hardware with an RGB-Depth Camera and Light Detection and Ranging Sensor (LiDAR). RGB-D Cameras are a specific type of depth sensing devices that are able to augment the conventional image with depth information on a per-pixel basis. A LiDAR emits pulsed light waves into the surrounding environment which then bounce off nearby obstacles and returns to the sensor. The device then calculates the distance using the time taken by each pulse to return. This process when repeated millions of times per second creates a precise 3-D map of the surrounding environment. Both RGB-D Cameras and LiDAR are crucial for developing a fully autonomous system [16].

Smart Mobile Application for the users to provide a more reliable, personalized and convenient experience. The mobile application will include the following features –
   a. Book Issue History
   b. Complete Library Book List
   c. Due Date Notification
   d. Submission/Re-Issue Confirmation with Real Time Notification
   e. Live Availability of a vacant Bot

These features will ensure a better User Interface and Experience (UI & UX)

Lastly, we aim to upgrade the overall UX by adding a Thin Film Transistor (TFT) Touch Screen. The TFT Display will completely remove the need for a Laptop Screen during submission adding to the robustness of the project.

## VI. Conclusion

In this paper, we present a solution to the problem of students being forgetful in returning issued library books through an IoT based Book Robot. The robot can be present in the student's hostels for submission or re-issuing of books. The Pi Camera will verify the student's QR Code plus Book Barcode and update the database. The bot can then be navigated back to the librarian. The bot is provided with a GPS Module, Distance Sensors and a live video stream through Pi Camera to assist the controller. Through this work, we have tried to present a solution to the problems of students with overdue books, how a robot can help them with better time management and forgetfulness. This can not only save students from paying fines due to late submission but also help the library management system by automating a major part of it. The simple technology behind the robot not only reduces human intervention but also makes it more reliable.